\documentclass[aps,prd,preprintnumbers,showpacs]{revtex4}
\usepackage[dvips]{graphicx}
\begin{document}

%
%

\preprint{Nisho-3-2009}
\title{Pair Creation in Electric Flux Tube and Chiral Anomaly}
\author{Aiichi Iwazaki}
\address{International Economics and Politics, Nishogakusha University,\\ 
Ohi Kashiwa Chiba  277-8585, Japan.}   
\date{Aug. 31, 2009}
\begin{abstract}
Using chiral anomaly,
we discuss the pair creation of massless fermions 
under the effect of magnetic field $\vec{B}$ when
an electric flux tube $\vec{E}$
parallel to $\vec{B}$ is switched on.
The tube is axial symmetric and 
infinitely long.
In the limit
$B\gg E$,
we can analytically obtain the spatial and temporal behaviors of 
the number density of the fermions, the 
azimuthal magnetic field 
generated by the fermions etc..
We find that the life time $t_c$ of the electric field is shorter as the width of the tube is narrower. 
Applying it to the glasma in high-energy heavy-ion collisions,
we find that color electric field decays fast such
as $t_c\simeq Q_s^{-1}$ with saturation momentum $Q_s$.
\end{abstract}
\hspace*{0.3cm}
\hspace*{1cm}

\maketitle

Pair creation of charged particles in a classical strong electric field 
has been extensively discussed\cite{tanji} for long time since the
discovery of Klain paradox,
although such phenomena have not yet been observed. 
The particle creation is known as Schwinger mechanism\cite{schwinger}.
Here, we are addressed with the problem in a quite different way from
previous ones\cite{HE,schwinger,can} and apply our results to 
color gauge fields (glasma) produced in high-energy heavy-ion collisions.

The Schwinger mechanism
have been discussed under spatially homegeneous or inhomogeneous 
( and time dependent or independent ) back ground electric fields
with or without the effects of background magnetic fields.  
It has been treated in various formalisms, i.e. proper time method originally used by
Schwinger, Heisenberg-Euler effective
action formalism\cite{HE}, or canonical formalism\cite{can}.
But only in a few papers 
back reactions of the produced particles have been discussed
by taking account of the effects of background magnetic fields. 
In particular, there are no such papers in which a magnetic field induced by the electric current
of the particles is analyzed under the background magnetic field. 
On the other hand,
the Schwinger mechanism in such a circumstance
is quite relevant to the realistic cases such as the glasma, neutron stars etc.
Thus, it is an important issue to analyze the effects of the back reaction
under the background magnetic fields.

The glasma
have recently received much attention. 
The gauge fields of the glasma are classical color electric and magnetic fields
produced as initial states of gluons in high energy heavy ion collisions.
We speculate that
the decay of the glasma leads to thermalized quark gluon plasma (QGP)
observed in recent RHIC experiments. 
Schwinger mechanism
is a possible one for the decay of the color electric field.
Indeed, such a mechanism was phenomenologically discussed\cite{iwazaki} about thirty years ago.  
Although the presence of the color electric field in the discussions 
was simply assumed, it has recently been confirmed on the basis of a fairly reliable
effective theory of QCD at high energy, that is, a model of color glass condensate (CGC)\cite{colorglass}.

According to a model of the CGC, 
the color electric and magnetic fields are homogeneous in the longitudinal
direction and inhomogeneous in the transverse directions;
we define that the longitudinal direction is parallel to the collision axis.
Hence, we may think that
they form color electric and magnetic flux tubes extending in the longitudinal direction. 
In our previous papers\cite{iwa,itakura,hii}, 
we have discussed a mechanism for the decay of the color magnetic field
and have reproduced the instability of the glasma
observed in numerical simulations\cite{venugopalan,berges}.

In this paper we discuss the decay of an electric flux tube under background magnetic field
by taking account of the back reactions. 
The distinctive point in our paper is to use chiral anomaly 
without addressing the explicit forms of fermion's wavefunctions.
Simply using the anomaly 
we can analytically obtain the number density of the fermions, 
the life time of the electric field, etc., all of
which include the back reaction of the produced fermions in the electric field. 
Furthermore, we can analytically obtain an azimuthal magnetic field around the flux tube generated by the fermions. 
This is a remarkable feature in the use of chiral anomaly. 
However, the use of the chiral anomaly is limited only 
for the system with massless fermions under
the presence of both electric and magnetic fields
in the limit $B\gg E$.

\vspace*{0.2cm}
Hereafter, we mainly discuss the pair creation of massless electrons and positrons in QED.
The generalization to QCD would be straightforward.
Before discussing the pair creation in an electric flux tube,
we first explain how we use
chiral anomaly in a simple model of a homogeneous electric field.
We assume tentatively  
that the fields $\vec{B}$ and $\vec{E}$ are spatially homogeneous and parallel (or antiparallel) with each other.
They are oriented in the $z$ direction; 
$\vec{B}=(0,0,B)$ with $B>0$ and $\vec{E}=(0,0,E)$.
Then, the energies of electrons with charge $-e<0$ and positrons with charge $e>0$ 
under the magnetic field $B=|\vec{B}|$, but with $\vec{E}=0$ are given by

\begin{equation}
\label{1}
E_N=\sqrt{p_z^2+2NeB} \quad \mbox{(parallel)} \quad \mbox{and} \quad E_N=\sqrt{p_z^2+eB(2N+1)} \quad \mbox{(antiparallel)},
\end{equation}
where $p_z$ denotes momentum in the $z$ direction and 
integer $N\ge 0$ does Landau level. The term of
``parallel" (``antiparallel") implies magnetic moment parallel (antiparallel) to $\vec{B}$. 
The magnetic moment of electrons (positrons) is antiparallel (parallel) to their spin.
Thus,
electrons (positrons) with spin antiparallel (parallel) to $\vec{B}$ can have zero energy states
in the lowest Landau level; their energy spectrum is given by $E_{N=0}=|p_z|\ge 0$. 
They are ``massless" states.
On the other hand, the other states cannot be zero energy states; their energy spectra are given
by $\sqrt{p_z^2+eBM}\ge eBM\ge eB$ with positive integer $M$.
They are ``massive" states whose masses increase with $B$.  

When the electric field is switched on,
the pair creation of electrons and positrons with the energies $E_N$ begin to
occur.
However, it is not probable that any states are produced 
with an equal production rate. Indeed, high energy states are
hard to be produced while low energy states are easy to be produced.
Thus,
the ``massive" states are hard to be produced by weak electric field $|E| \ll B$.
In particular, they cannot be produced in the limit of $B \to \infty $.
Only products in the limit are the pairs of electrons and positrons in the ``massless" states.
We assume such a limit $B \gg |E|$ in our discussion below.

Note that owing to the Fermi statistics
only fermions with $p_z=0$ are produced,
since the other states with $p_z\neq 0$ have already been occupied.
After the production of the fermions with $p_z=0$, their momenta
increase with time, $p_z(t)=\pm e\int^t dt' E$, owing to the acceleration by the electric field.
Hence we obtain the momentum distribution\cite{tanji2} $\tilde{n}(p_z)\propto \theta(p_F(t)-p_z)\theta(p_z)$ 
for positrons with charge $e>0$ and $\tilde{n}(p_z)\propto \theta(p_F(t)+p_z)\theta(-p_z)$
for electrons. 
The Fermi momentum $p_F$ is given by $p_F(t)=e\int_0^tdt'E(t')$.
Here we have assumed that the electric field is parallel to $\vec{B}$
and is switched on at $t=0$.  

We note that electrons move to the direction antiparallel to $\vec{E}$ while
positrons move to the direction parallel to $\vec{E}$.
Therefore, both electrons and positrons created by the electric field have right handed helicity
when $\vec{E}$ being parallel $\vec{B}$, 
while they have left handed helicity when $\vec{E}$ being antiparallel $\vec{B}$. 
The helicity of the particles never change in the course of their propagation.


The key point in our discussion is to use the equation of chiral anomaly,

\begin{equation}
\label{chiral}
 \partial_t (n_R-n_L)=\frac{e^2}{4\pi^2}\vec{E}(t)\vec{B}
\end{equation}
where $n_R$ ( $n_L$ ) denotes number density of right ( left ) handed chiral fermions; 
$n_{R,L}=\langle\bar{\Psi}\gamma_0(1\pm \gamma_5)\Psi\rangle/2$
in which the expectation value is taken by using a state of electrons and positrons created
in the electric field. Here, we have used spatial homogeneity of the chiral current $\vec{j}_5$, that is, $\rm{div}\vec{j}_5=0$.
Since we have assumed sufficiently strong magnetic field $B\gg E$, 
the back reaction of the produced particles on $B$ is negligible.
( Although the magnetic moments of the fermions partially screen the back ground field $B$, 
the effect is negligible in the limit of $B\gg E$. )

In accordance with the chiral anomaly, 
the rate of chirality change is given by the product of $\vec{E}$ and $\vec{B}$.
It is important to note that 
the chirality is identical to the helicity in the case of massless fermions.
Thus, the anomaly equation describes the rate of helicity change. 
As the helicity of the particles does not change in the course of their propagation,
the equation describes the rate of the particle production or annihiration.
Since
only particles with right ( left ) handed helicity
are produced in the limit $B\gg E$ when $\vec{E}$ is 
parallel to $\vec{B}$ ($\vec{E}$ being antiparallel to $\vec{B}$ ), 
$n_R=2n$ and $n_L=0$ with the number density $n$ of electrons when $\vec{E}$ parallel to $\vec{B}$, 
while $n_R=0$ and $n_L=2n$ 
when $\vec{E}$ antiparallel to $\vec{B}$.
Thus, we find that the pair production
is determined by the anomaly equation
$2\partial_t n(t)=e^2E(t)B/4\pi^2$ with $E(t)>0$.

Similarly, the chiral anomaly describes the pair annihilation $\partial_t n=e^2E(t)B/4\pi^2<0$ when $E(t)<0$.
The annihilation may occur subsequent to the pair creation.
Indeed, after the particle production the electric field $E>0$ loses its energy and vanishes 
because the field accelerates the charged particles.
Then, the field changes its sign, i.e. $E<0$ since a Maxwell equation $\partial_tE(t)=-J$ 
with non vanishing electric current $J=2en(t)$ 
of the particles,
dictates that $E$ should not remain zero ( see later. ) 
Negative $E$ in the anomaly equation implies the decrease of the number density $n$ with time, 
which is caused by pair annihilation.

In order to discuss back reactions of the particles, we need to consider an energy
conservation between the
energies $\epsilon(t)$ of the particles and the energy of the field,

\begin{equation}
\label{3}
\partial_t\Bigl(\epsilon(t) +\frac{1}{2}E^2(t)\Bigr)=0
\end{equation} 
where we have neglected the contribution of magnetic field induced by the electric current of electrons and positrons.
This is allowed in the system with spatially homogeneous fields $E$ and $B$.
Later we take into account the contribution when we discuss an electric flux tube.

The energy density $\epsilon(t)$ of the particles are given
such as $\epsilon(t)=2n(t)p_F(t)/2=n(t)p_F(t)$.
This originates from the fact that the momentum distribution $\tilde{n}(p_z)$ ( $n(t)\equiv \pm\int_0^{\pm\infty} dp_z \tilde{n}(p_z)$ )
is given by $\tilde{n}(p_z)=n_0\theta (p_F(t)-|p_z|)\theta(|p_z|)$ as we have discussed.
For example, the energy density of electrons is given such that $\int_{-\infty}^0 dp_z \tilde{n}(p_z)|p_z|=np_F/2$.

Now we have three equations to solve for obtaining the pair production rate $\partial_t n(t)$, etc.,

\begin{equation}
\label{tot}
2\partial_t n(t)=\frac{e^2}{4\pi^2}E(t)B, \quad \partial_t\Bigl(\epsilon(t) +\frac{1}{2}E^2(t)\Bigr)=0, \quad \mbox{and}
\quad \epsilon(t)=n(t)p_F(t)
\end{equation}
with $p_F(t)=\int_0^t dt' eE(t')$.
It is easy to solve the eq(\ref{tot}) with initial conditions $E(t=0)=E_0>0$ and $n(t=0)=0$ by
assuming magnetic field $B$ independent on $t$,

\begin{equation}
\label{result}
E(t)=E_0\cos(\sqrt{\frac{\alpha eB}{\pi}}\,t) \quad \mbox{and} \quad 
n(t)=\frac{\alpha E_0B|\sin(\sqrt{\frac{\alpha eB}{\pi}}\,t)|}{2\pi \sqrt{\frac{\alpha eB}{\pi}}}
\end{equation}
with $\alpha=e^2/4\pi$.

\vspace*{0.2cm}
In Fig.\,1 we have shown the behaviors $E(t)=E_0\cos(\pi \,t/2t_c)$ and $n(t)=\frac{\alpha E_0 Bt_c}{\pi^2}|\sin(\pi\,t/2t_c)|$ 
with $E_0=1$, $t_c=1$ and $\alpha\simeq 1/137$, where $t_c \equiv \frac{\pi}{2}(\sqrt{\frac{\alpha eB}{\pi}})^{-1}$.
The formula holds rigorously in the limit of $B\gg E_0$ where only right handed or left handed particles are produced.
The oscillating features of $E(t)$ and $n(t)$ have agreed with those obtained in the previous papers\cite{tanji},
in which the wavefunctions of electrons under the effects of the electric and magnetic fields
have been explicitly used.
We also note that the screening of the magnetic field $B$ by the magnetic moments 
of the fermions is negligible in the limit $B\gg E$ since 
the number density $n$ of the fermions is proportional to $\sqrt{B}$.

Here we wish to explain why the number density $n$ and the electric field $E$ oscillate.
When the electric field $E>0$ is switched on at $t=0$, 
the pair production of the particles with {\bf right handed helicity} begins to occur 
and their number density increases.
Owing to the acceleration of the particles by the electric field, 
the Fermi momentum $p_F(t)=\int_0^t dt'eE(t')$ and their energies increases. On the other hand,
the energy of the electric field gradually decreases and vanishes at $t=t_c$.
At the time the number density takes the maximum value. Then, the electric field changes its direction
( $E<0$ for $t>t_c$ ) and becomes strong with time, as discussed before. Accordingly, the number density decreases after $t=t_c$ 
since $2\partial_t n(t)=\frac{e^2}{4\pi^2}E(t)B <0$. 
This decrease is caused by the pair annihilation of electrons and positrons. Since the direction of the electric field changes
after $t_c$, the direction of the particle acceleration also change so that the Fermi momentum $p_F(t)$ begins to decrease.
Consequently, electrons and positrons are moved to overlap with each other so that the pair annihilation
may occur to make the number density decrease. Eventually, all of the particles with {\bf right handed helicity} vanishes at $t=2t_c$. 
At the time the strength of the field $E(t=2t_c)=-E_0<0$ becomes maximum. Since $E$ is negative and $n=0$ at $t=2t_c$, 
a new pair creation of the particles with {\bf left handed helicity} begins to occur
after $t\geq 2t_c$. 
Their number density increases in accordance with the chiral anomaly,  
$2\partial_t n(t)=-\frac{e^2}{4\pi^2}E(t)B >0$. 
Therefore, the oscillation of $E(t)$ and positiveness of $n(t)$ as shown in Fig.1 arises. 

Here we wish to make a comment why the electric field increases after $t=t_c$.
Since all the positrons ( electrons ) move to the direction of $\vec{B}$ ( $-\vec{B}$ ) at $t=t_c$, the electric field
anti-parallel to their electric current becomes strong by gaining its energy from the particles, once
such an electric field is present even if it is small. on the other hand, the electric field parallel to the current
vanishes losing its energy.
Therefore, the electric field increases with its direction anti-parallel to $\vec{B}$ when
$t>t_c$.

In the above discussion, we have assumed that the mass $m_e$ of electrons and positrons is vanishingly small.
The assumption may hold as far as $E>m_e^2$. 
Thus, when the electric field becomes small such as $E<m_e^2$ by losing the energy, 
the anomaly in eq(\ref{chiral}) cannot be used. Actually the pair creation can hardly arise in such a case. However, 
the field loses its energy due to the acceleration of the particles. Thus, it vanishes ( $E=0$ )
and increases with the opposite direction ( $E<0$ ) in acoordance with the Maxwell equation as explained above. 
Once it arises with its direction opposite to that of the electric current, it gains the energy.
Consequently, the field can becomes strong such as $E>m_e^2$
so that the assumption of the masslessness approximately holds again.
Although the detail in the real behavior of $E$ is different with that in Fig.1 when $E$ is small,
the feature of the oscillation we found still remains.

\begin{figure}[t]
\begin{minipage}{.47\textwidth}
\includegraphics[width=6.9cm,clip]{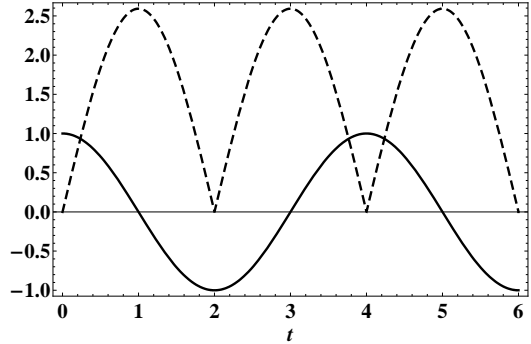}
\label{fig:growth-rate3}
\caption{electric field $E(t)$ (solid line) and number density $n(t)$ (dashed line)}
\end{minipage}
\end{figure} 

It is important to note that although the formula of the chiral anomaly involves all of the quantum effects, 
the mechanism why the number density increases or decreases
is not explicit in the formula. However,
we understand that the increase or decrease of the number density originates with the pair creation or annihilation.
This is because the results shown in this paper have agreed with those obtained by
calculations\cite{tanji} explicitly including the quantum effects of the pair creation or annihilation.


\vspace*{0.2cm}
Now we discuss the pair production in an axial symmetric electric flux tube, namely the electric field
extending infinitely in the z direction, but has finite width $R$ in the transverse directions.
In this case the quantities of $E$, $n$ and $p_F$ depend on the
cylindrical coordinate $r=\sqrt{x^2+y^2}$ and time coordinate $t$. 
( Since the wavefunctions in the Lowest Landau level are given by $Z^m\exp(-eB|Z|^2/4)$ with $Z\equiv x+iy$ and integer $m\ge 0$,
the charged particles can be located within the flux tube for sufficiently large $B$ such as $R\gg 1/\sqrt{eB}$. )
Thus, the formula of the chiral anomaly takes
the same form as the one in eq(\ref{tot}), although the electric field $E(r,t)$ and the number density $n(r,t)$
depend on $r$ and $t$. ( In the equation of the chiral anomaly,
we may take $div\vec{j}_5=\partial_zj_z^5+\sum_{i=x,y}\partial_ij_i^5=0$,
because of the homogeneity in the z direction and the fact that 
$\sum_{i=x,y}\partial_ij_i^5=\sum_{i=x,y}\partial_i \langle\bar{\Psi}\gamma_5\gamma_i\Psi\rangle=0$ for
the ``massless" states in the Lowest Landau level. )
Similarly, the energy density $\epsilon$ takes the form $\epsilon(r,t)=n(r,t)p_F(r,t)$ as eq(\ref{tot}),
in which the Fermi momentum is given by $p_F(r,t)=\int_0^{t}dt'eE(r,t')$.
We also take into account Maxwell equations such as $\partial_t B_{\theta}(r,t)=\partial_r E(r,t)$ and
$\partial_t E(r,t)=\frac{\partial_r(rB_{\theta}(r,t))}{r}-J(r,t)$ where $J(r,t)$ represents the current of electrons and positrons
flowing in the $z$ direction. $B_{\theta}(r,t)$ denotes azimuthal magnetic field generated by the current.
Therefore, we have the following equations to solve,

\begin{equation}
\label{6}
2\partial_t n(r,t)=\frac{e^2}{4\pi^2}E(r,t)B, \quad \epsilon(t)=n(t)p_F(t), 
\end{equation}
and Maxwell equations,
\begin{equation}
\label{max}
\partial_t B_{\theta}(r,t)=\partial_r E(r,t),
\quad \partial_t E(r,t)=\frac{\partial_r(rB_{\theta}(r,t))}{r}-J(r,t).
\end{equation}
where $B$ is assumed independent of $r$ and $t$ because back reactions of the fermions on $B$ are small
in the limit $B\gg E$.

In order to solve the equations we must represent the current $J(r,t)$ in terms of $n(r,t)$ and $p_F(r,t)$.
It can be done by imposing the condition of energy conservation,

\begin{equation}
\label{cond}
\partial_t\int d^3x \Bigl(\frac{E^2(r,t)+B_{\theta}^2(r,t)}{2}+\epsilon(r,t)\Bigr)=\int d^3x \Bigl(E(r,t)\partial_t E(r,t)
+B_{\theta}(r,t)\partial_tB_{\theta}(r,t)+\partial_t\epsilon(r,t)\Bigr)=0.
\end{equation}
The condition implies that the energy of the electric field is transmitted into the energies of
the particles and the azimuthal magnetic field $B_{\theta}$.

Using the equations in eq(\ref{6}) and eq(\ref{max}), we rewrite the condition in eq(\ref{cond}) such that

\begin{equation}
\int d^3x \Bigl(E(\partial_tE-\frac{\partial_r(rB_{\theta})}{r})+\partial_t\epsilon \Bigr)=
\int d^3x(-JE+\partial_t\epsilon)=\int d^3x E(en+\frac{e^2}{8\pi^2}B p_F-J)=0,
\end{equation}
where we have performed the partial integration in $r$.
The energy conservation must hold for any initial conditions of $E(r,t=0)=E_0(r)$.
Therefore, we obtain 

\begin{equation}
\label{J}
J(r,t)=en(r,t)+\frac{e^2}{8\pi^2}p_F(r,t)B=2en(r,t),
\end{equation}
since $\frac{e^2}{8\pi^2}p_F(r,t)B=\frac{e^2}{8\pi^2}\int_0^t dt' eE(r,t')B=en(r,t)$ with $n(r,t=0)=0$.
It states that both of electrons with charge density $-en$ and positrons with $en$ 
constitute the electric current $J$ since the velocity of electrons is $-1$, while that of positrons is $+1$
in the unit of light velocity $c=1$. 
This formula of $J=2en$ has been used in the discussion just above eq(\ref{3}).

Using $J(r,t)$ in eq(\ref{J}) as well as equations in eq(\ref{6}) and eq(\ref{max}), we derive the equation of motion of the electric field,

\begin{equation}
\label{E}
\partial_t^2E(r,t)=(\partial_r^2+\frac{1}{r}\partial_r-\frac{e^3B}{4\pi^2})E(r,t),
\end{equation}
where we have used the formula $\partial_tJ(r,t)=\frac{e^3}{4\pi^2}BE(r,t)$.

We solve the equation(\ref{E}) with the initial conditions $n(r,t=0)=0$, $B_{\theta}(r,t=0)=0$ and $E(r,t=0)=E_0\exp(-r^2/R^2)$.
The initial conditions implies that the electric flux tube with radius $R$ is switched on at $t=0$
when any particles and magnetic field $B_{\theta}$ are absent. 
Using the initial conditions $E(r,t=0)=E_0\exp(-r^2/R^2)$ and $\partial_t E(r,t=0)=0$ derived from Maxwell equations (\ref{max}),
we first obtain the solution $E$ in eq(\ref{E}), and then we derive $n$ and $B_{\theta}$, 

\begin{eqnarray}
\label{sol1}
E(r,t)&=&\frac{E_0R^2}{2}\int_0^{\infty}kdk \cos(t\sqrt{k^2+m^2})J_0(kr)\exp(-k^2R^2/4), \\
\label{sol2}
n(r,t)&=&\frac{e^2BE_0R^2}{16\pi^2}\Bigl|\int_0^{\infty} kdk \frac{\sin(t\sqrt{k^2+m^2})}{\sqrt{k^2+m^2}}J_0(kr)\exp(-k^2R^2/4)\Bigr|, \\
\label{sol3}
B_{\theta}(r,t)&=&-\frac{E_0R^2}{2}\int_0^{\infty} k^2dk \frac{\sin(t\sqrt{k^2+m^2})}{\sqrt{k^2+m^2}}J_1(kr)\exp(-k^2R^2/4),
\end{eqnarray}
with $m^2\equiv\frac{e^3}{4\pi^2}B$,
where $J_i(kr)$ denotes Bessel functions. 
Since the integration over $k$ is dominated by small $k$ in the limit $R\to \infty$,
we find that the solutions in eq(\ref{sol1}) $\sim$ eq(\ref{sol3}) 
are reduced to the previous solutions in eq(\ref{result})
in the limit.
We note that the effect of $B_{\theta}$ on the particles is negligible compared with the background 
magnetic field $B$ since $B_{\theta} \propto 1/\sqrt{B}$ in the limit $B\to \infty $. 

We have shown the temporal and spatial behaviors of electric field $E(r,t)$ in Fig.2, number density $n(r,t)$ in Fig.3,
and azimuthal magnetic field $B_{\theta}$ in Fig.4, respectively, where the parameters $m=1$, $E_0R^2/2=1$, $e^2/8\pi^2=1$ and $R=1$ have been used;
the scale $r=10$ in the figures corresponds to the width $R$ of the electric flux tube. Small and large dots represent the behaviors at $t=0.2$ and $t=0.6$, respectively. 
We find that the stronger electric field produces the particles more. It leads to the fact 
that the number density of the particles is larger as $r$ is smaller. Thus,
the electric field $E(r,t)$ vanishes faster as $r$ is smaller since $E(r,t=0)$ is stronger as $r$ is smaller.
We also find that the life time of the electric field with finite $R$ is
shorter than that of the field with $R=\infty$. ( The life time may be defined roughly as $E(r=R,t=t_c)=0$,
although the field oscillates with time. )
This is because typical momentum $k$ dominating in the above integration is on the order of $R^{-1}$. 
Thus, the life time is approximately given by $t_c\simeq (\sqrt{m^2+R^{-2}})^{-1}$.

\begin{figure}[t]
\begin{minipage}{.47\textwidth}
\includegraphics[width=7.1cm,clip]{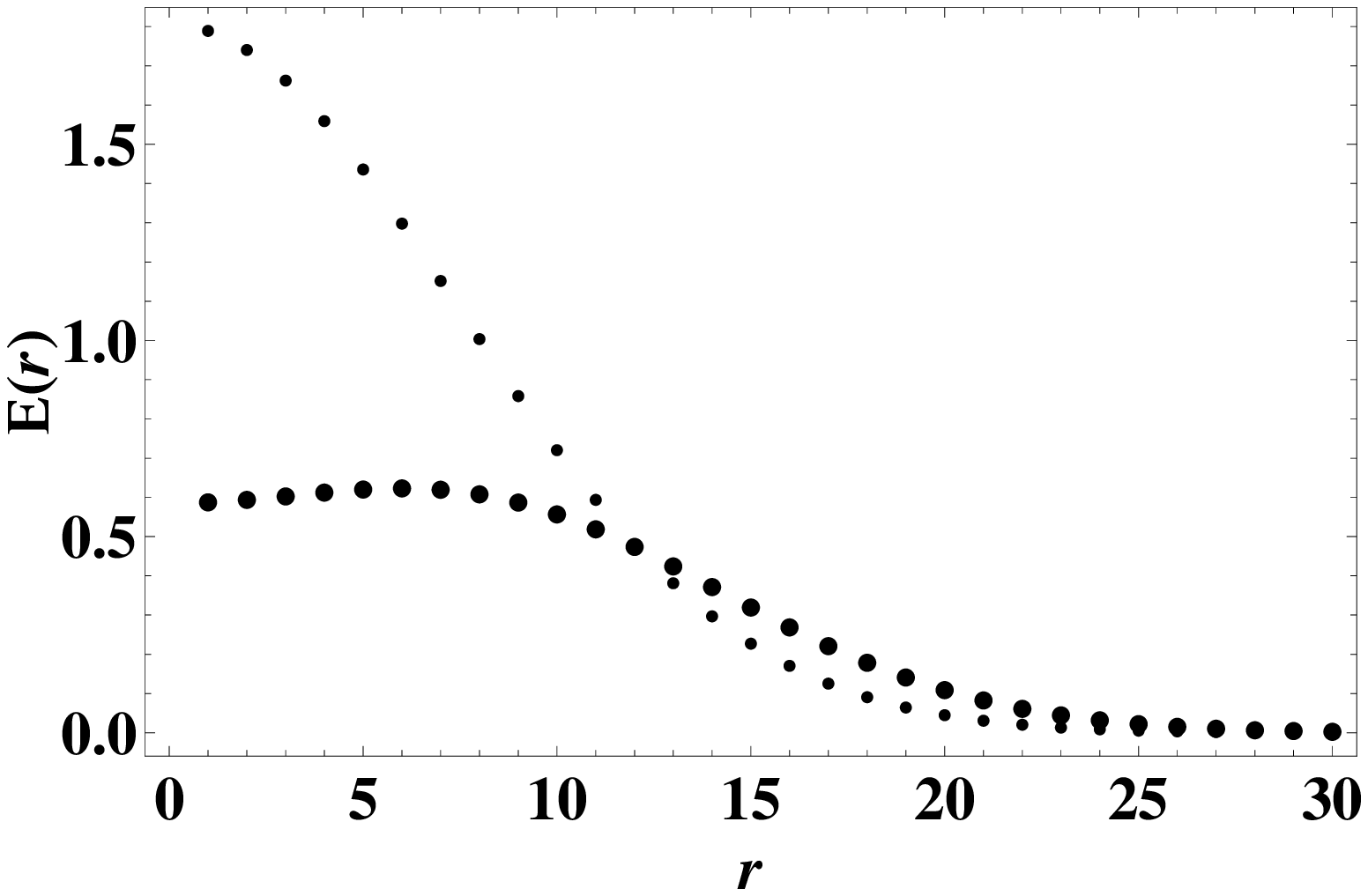}
\label{fig:E}
\caption{electric field $E(r)$ at $t=0.2$ (small dots) \\ and $E(r)$ at $t=0.6$ (large dots)}
\end{minipage}
\hfill
\begin{minipage}{.47\textwidth}
\includegraphics[width=7.1cm,clip]{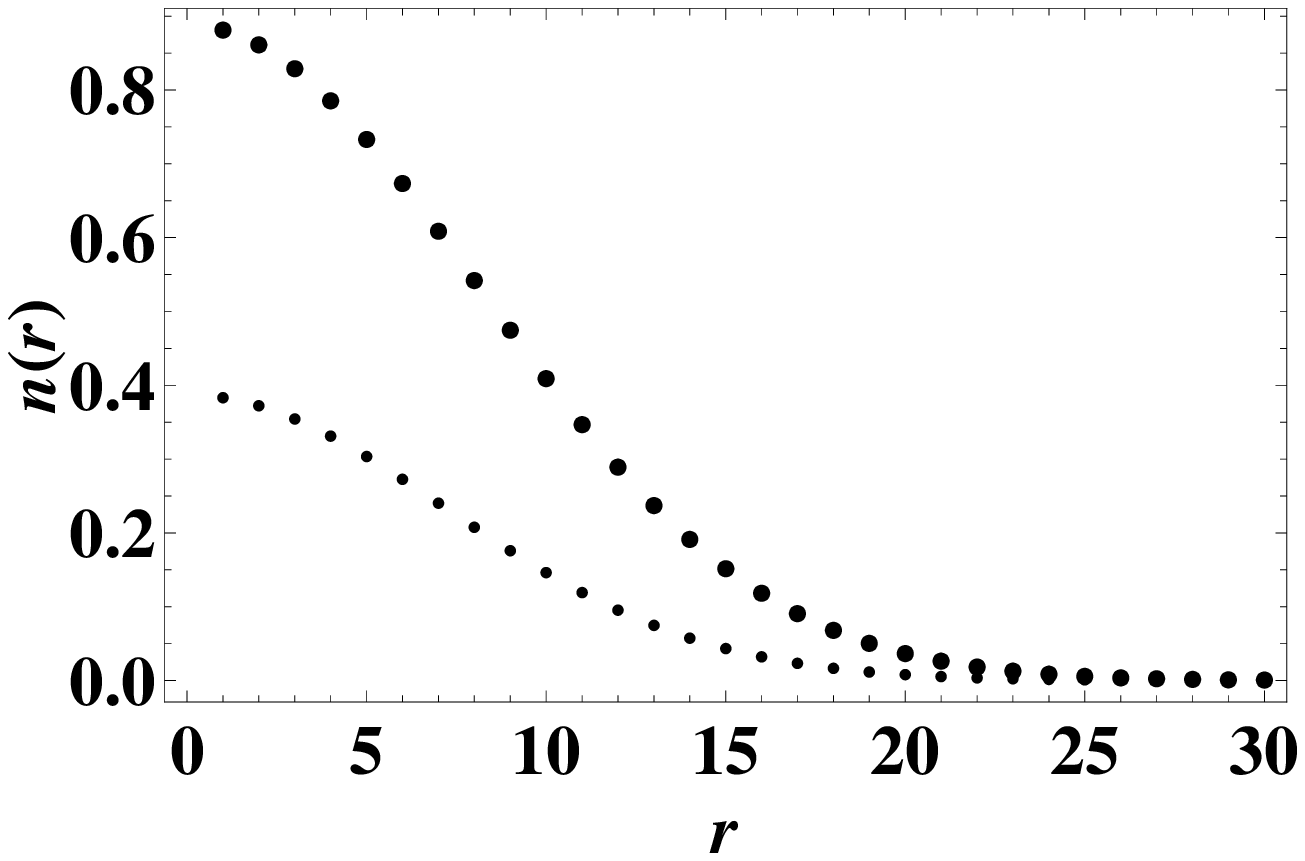}
\label{fig:n}
\caption{number density $n(r)$ at $t=0.2$ (small dots) \\ and $n(r)$ at $t=0.6$ (large dots)}
\end{minipage}
\end{figure}

\begin{figure}[t]
\begin{minipage}{.47\textwidth}
\includegraphics[width=7.1cm,clip]{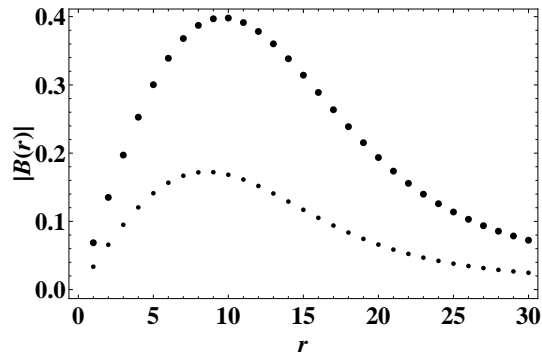}
\label{fig:B}
\caption{azimuthal magnetic field $B_{\theta}(r)$ at $t=0.2$ \\(small dots) and $B_{\theta}(r)$ at $t=0.6$ (large dots)}
\end{minipage}
\end{figure}

The quantities $E(r,t)$, $n(r,t)$, and $B_{\theta}(r,t)$ oscillate with time
owing to the similar reason explained above.
In particular, electric field $E(r,t)$ never disappear by completely losing its energy.
This originates with the fact that the produced particles are almost free
and do not interact with externals e.g. external heat bath so that their energies never dissipate.
On the other hand, when we take into account their interactions with external heat bath and 
use fermionic momentum distributions at finite temperature,
the electric field would decay and disappear
because the current $J$ would vanish owing to the dissipation of the particle's momenta
in the heat bath. 



\vspace*{0.3cm}

Finally we wish to make a comment on the decay of the glasma produced in high-energy heavy-ion collisions.
The color electric and magnetic fields generated in the early stage of the collisions form flux tubes
and their field strenghts are almost identical to each other.
Although we have assumed that the gauge fields are Abelian 
and magnetic field $B$ is not a tube-like but homogeneous and $B \gg E$,
we may approximately apply our results to the analysis of the decay of the color electric field. 
In particular we use the result that the life time $t_c$ of electric field is approximately given by $(\sqrt{R^{-2}+m^2})^{-1}$.
In the glasma, $m^2$ is of order $\alpha_s gB$ where $gB \simeq Q_s^2$ with saturation momentum $Q_s$ 
( $\alpha_s= g^2/4\pi \simeq 1/4\pi$ with gauge coupling constant $g\simeq 1$ ),
while the width of the flux tube $R$ is given by $Q_s^{-1}$. 
Hence, we find that the color electric field
decays rapidly such as $t_c\simeq Q_s^{-1} \simeq 0.1\,\rm{fm/c} \sim 0.2\,\rm{fm/c}$
for $Q_s=1\,\rm{GeV}\sim 2\,\rm{GeV}$. The smallness of the width is the main
origin of the rapid decay. 
Although our result does not include expanding effects of the glasma,
the result suggests that the color electric field decays sufficiently fast
as required phenomenologically\cite{hirano}. In addition,
the result is consistent with our assumption of
the independence of $B$ on time. 
Indeed, $B$ decays very slowly\cite{venugopalan,lappi} 
and does not vary so much at least within the period of $Q_s^{-1}$.

\vspace*{0.3cm}
To summarize, using the chiral anomaly we have 
discussed the pair creation of massless charged fermions in an axial symmetric electric flux tube
under the effect of strong magnetic field.
We have analytically obtained
the spatial and temporal behaviors of their number density, the electric field
and the azimuthal magnetic field induced by the current of the fermions.
Our calculations have been performed 
without addressing explicit forms of particle's wavefunctions.
These quantities of the free massless fermions are exact in the limit of $B\gg E$. 
Applying our results to the glasma in high energy heavy-ion collisions, 
we have found that the color electric field decay sufficiently fast to be consistent
with QGP phenomenology.

\hspace*{1cm}

The author
express thanks to Drs. H. Fujii of University of Tokyo and Dr. K. Itakura
of KEK for their useful discussion and comments. He also thanks to Dr. Tanji
of University of Tokyo for his numerical result of the momentum distribution when $m_e\to 0$.


\end{document}